\documentclass[prd,reprint,amsmath,amssymb,floatfix,aps,longbibliography,superscriptaddress,nofootinbib]{revtex4-2}

\usepackage{amsmath,graphicx,amssymb,xcolor,booktabs,braket,multirow}
\usepackage[colorlinks=true, allcolors=blue]{hyperref}
\usepackage{soul}

\setstcolor{red}

\begin{document}

\title{Baryon number violating rate as a function of the proton-proton collision energy}

\author{Yu-Cheng Qiu}
\email{ethanqiu@sjtu.edu.cn}
\affiliation{Tsung-Dao Lee Institute and School of Physics and Astronomy, \\
Shanghai Jiao Tong University, 520 Shengrong Road, Shanghai, 201210, China}

\author{S. -H. Henry Tye}
\email{tye.henry@gmail.com}
\affiliation{Department of Physics,\\
Hong Kong University of Science and Technology, Hong Kong S.A.R., China}
\affiliation{Department of Physics,\\
Cornell University, Ithaca, New York 14853, USA}

\date{December 10, 2023}

\begin{abstract}
The baryon-number violation (BV) happens in the standard electroweak model. According to the Bloch-wave picture, the BV event rate shall be significantly enhanced when the proton-proton collision center of mass (COM) energy goes beyond the sphaleron barrier height $E_{\rm sph}\simeq 9.0\,{\rm TeV}$. Here we compare the BV event rates at different COM energies, using the Bloch-wave band structure and the CT18 parton distribution function data, with the phase space suppression factor included. As an example, the BV cross section at 25 TeV is 4 orders of magnitude bigger than its cross section at 13 TeV. The probability of detection is further enhanced at higher energies since an event at higher energy will produce on average more same sign charged leptons. 

\end{abstract}

\maketitle

\section{Introduction}\label{sec:intro}

Matter-antimatter asymmetry is an important mystery in our Universe. The baryon-number violation (BV) via the instanton~\cite{Belavin:1975fg} in the standard electroweak model observed by 't Hooft~\cite{tHooft:1976rip,tHooft:1976snw} provides a crucial avenue to understanding baryogenesis. Therefore, observing (confirming) such BV in the laboratory will be immensely valuable.

The underlying physics of the BV process can be reduced to a simple quantum mechanical system. With the Chern-Simons (CS) number $Q$ (or $n=m_W Q /\pi$) as the coordinate, one obtains the one-dimensional time-independent Schr\"{o}dinger equation, with mass $m\simeq 17$ TeV~\cite{Tye:2015tva}:
\begin{equation}\label{eq:QMeq}
 \left[ - \frac{1}{2m}\frac{\partial ^2}{\partial Q^2} +V(Q) \right] \Psi(Q) = E\Psi(Q)\;, 
 \end{equation}
where the sphaleron potential $V(Q)$ is periodic, with minima at integer values of $n$ and maxima at $n+1/2$, with barrier height $E_{\rm sph}=9.0\,{\rm TeV}$~\cite{Manton:1983nd,Klinkhamer:1984di,Akiba:1988ay}. Although this Schr\"{o}dinger equation is well accepted, it is the interpretation of the underlying physics of $V(Q)$'s periodicity that needs clarification: whether the solution of this equation has a Bloch wave band structure or not.

Let us first consider the $SU(2)$ gauge theory without the fermions: in this case, all integer $n$ states are physically identical; that is, $n \to n \pm 1$ simply goes back to itself (though in a different gauge). So there is no band structure, as is the case in the QCD theory. This is analogous to a rigid pendulum rotating by $2 \pi$ via tunneling~\cite{Bachas:2016ffl}.

Once left-handed fermions couple to the electroweak $SU(2)$ gauge theory,
different $n$ state has different baryon (and lepton) numbers, so they are physically different: as we go from the $n$ to the $n + \Delta n$ state, baryon number changes by $3\Delta n$. As $Q$ runs from $-\infty$ to $+\infty$, a band structure emerges. Changing $Q$ is no longer exponentially suppressed within each band. For energies below the height of the sphaleron potential of $9.0$ TeV, band gaps dominate over the bandwidths, so the BV cross section $\sigma_{\rm BV}$ is still small. As $E$ increases, the bandwidths grow while the gaps between bands decrease.  Once the energy goes above $9.0$ TeV, bands take over, and the BV cross section is no longer exponentially suppressed. This is in contrast to the QCD theory which has no bands.

The Large Hadron Collider (LHC) at CERN ran at proton-proton collision energy $E_{pp}=13$ TeV and is presently running at $E_{pp}=13.6$ TeV. Since the quarks and gluons inside a proton share its energy, the quark-quark energy $E_{qq}$ is only a fraction of the total $E_{pp}$. It is important to see how $\sigma_{\rm BV}$ grows as $E_{pp}$ increases. This is a simple kinematic issue. Reference~\cite{Ellis:2016ast} has estimated the growth of $\sigma_{\rm BV}$ as a function of $E_{pp}$. Here we like to dwell into the estimate in more detail by taking the band structure fully into account as well as an additional phase space factor: even if $E_{qq} > 9.0$ TeV, not all energy goes to the BV process. That is, $E_{qq}$ has to be shared between baryon-number conserving (BC) scattering and BV scattering. In this note, we present for $E_{pp}$ above $13$ TeV, the ratio 
\begin{equation}
\eta (E_{pp}) = \frac{\sigma_{\rm BV} (E_{pp})}{\sigma_{\rm BV} (13 \,{\rm TeV})}\;.
\label{eq:eta}
\end{equation}
%As shown in Fig.~\ref{fig:sigma_s}, the BV cross section at $E_{pp}=25\,{\rm TeV}$ is $4$ orders-of-magnitude larger than that at $E_{pp}=13\,{\rm TeV}$.
%increasing to $E_{pp}=25\,{\rm TeV}$.

\begin{figure}
    \centering
    \includegraphics[width=7.5cm]{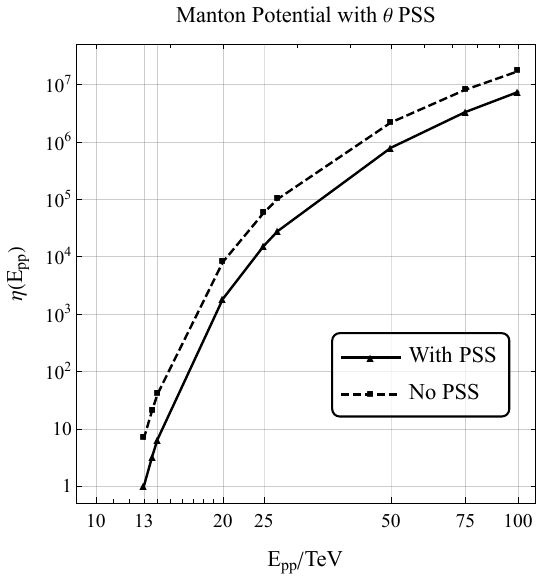}
    \caption{Solid curve is $\eta(E_{pp})=\sigma_{\rm BV} (E_{pp})/\sigma_{\rm BV} (13 \,{\rm TeV})$ with the $\theta$ phase space suppression (PSS) as a function of $E_{pp}$. $\theta$ stands for the parameter describing the energy budget for the BV process in total $E_{qq}$ as explained in Sec.~\ref{sec:theta}. The dashed curve is the $\eta(E_{pp})$ without the PSS, whose $\eta$ is normalized to the phase-space-suppressed $\sigma_{\rm BV}(13\,{\rm TeV})$ for comparison. Two different parametrizations of Sphaleron potential named Manton and AKY potentials give similar band structures (see Table~\ref{tab:bands}). They give almost identical (up to 2 significant digits) enhancement $\eta(E_{pp})$ here.  }
    \label{fig:sigma_s}
\end{figure}

A rough estimate assumes a cutoff model, which states that $\sigma_{\rm BV}$ is totally suppressed for $E_{qq} < 9.0$ TeV and completely unsuppressed for $E_{qq} \ge 9.0$ TeV. 
As an exercise, we first present an analytical evaluation of $\eta (E_{pp})$. However, as we shall see, this estimate is not accurate enough. Using the parton distribution function (PDF) for the valence quarks from the CTEQ program~\cite{Hou:2019efy}, the estimate for $\eta (E_{pp})$ agrees with that in Ref.~\cite{Ellis:2016ast}. Next, we take the band structure into account: $\sigma_{\rm BV}$ is completely unsuppressed for $E_{qq}$ inside a Bloch band and totally suppressed for $E_{qq}$ in a band gap. It turns out this result is close to the above simple estimate if we choose the critical $E_{qq}=9.1$ TeV instead of $9.0$ TeV.
However, even inside a Bloch band, not all $E_{qq}$ goes to BV scatterings; some energies flow to the baryon-conserving (BC) channel.

\begin{figure}
    \centering
    \includegraphics[width=7.5cm]{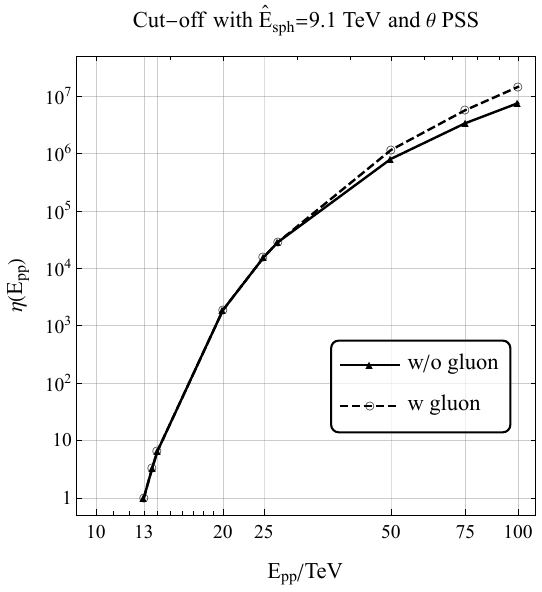}
    \caption{The solid curve is $\eta(E_{pp})$ without gluon contribution. The dashed curve is $\eta(E_{pp})$ with gluon $+$ quark scattering included for $50$--$100$ TeV. Here they are calculated under the cutoff model with the effective cutoff at ${\hat E}_{\rm sph}=9.1\,{\rm TeV}$ and $\theta$ phase space suppression.}
    \label{fig:eta_with_G}
\end{figure}

We also perform estimates on the BV cross section including this phase space suppression factor, again using parton distribution functions (PDFs) from the CTEQ program~\cite{Hou:2019efy}. Our representative final result is presented in Fig.~\ref{fig:sigma_s}. We see that $\sigma_{\rm BV}(25\, {\rm TeV})$ is 4 orders-of-magnitude bigger than $\sigma_{\rm BV}(13\, {\rm TeV})$. 
Including gluon $+$ quark scattering has little effect on the result as gluon PDF is rather soft, as shown in Fig.~\ref{fig:eta_with_G}.

There is another important effect that should come into play. Here 
$\eta(E_{pp})$ (\ref{eq:eta}) only compares $\sigma_{\rm BV}$ at different energies. Based on the analysis of Ref.~\cite{Qiu:2018wfb}, we expect that $\sigma_{\rm BV}(25\, {\rm TeV})$ will involve events with larger $\Delta n$ than $\sigma_{\rm BV}(13\, {\rm TeV})$. Although it is hard to estimate the enhancement of $\Delta n$ as one increases the energy, it is likely that the average $\langle \Delta n \rangle$ at 25 TeV is an order-of-magnitude bigger than the average $\langle \Delta n \rangle$ at 13 TeV. Since a single $|\Delta n|$ event can produce up to $3|\Delta n|$ same sign charged leptons, the probability of BV detection will be substantially enhanced beyond that coming only from an increase in $\eta (E_{pp})$.

\section{Estimate of $\eta(E_{pp})$}

%An estimate of the cross section as a function of the parton distribution functions and the pass (Bloch wave) bands will be very useful. 

%Let us formulate the problem here.
Consider proton-proton ($pp$) collisions. In the center of mass (COM) frame, the proton momenta are
$P_1=(E,0,0,E)$ and $P_2=(E,0,0,-E)$. where $s=4E^2$. So the quark-quark momentum is 
\begin{equation}
v=x_1P_1+x_2P_2=\left((x_1+x_2)E,0,0,(x_1-x_2)E\right)\;,
\label{eq:v}
\end{equation}
where $x_j$ is the fraction of momentum carried by quark $q_j$. The invariant energy carried by the quark-quark system is $v$, where
\begin{equation}
    v^2= x_1 x_2 s\;,
\end{equation}
%where the COM energy $E_{pp}=\sqrt{s}$.
%$=13\,{\rm TeV}$ before and is now running at $13.6\,{\rm TeV}$ but so far reluctant to go to $14\,{\rm TeV}$. Future collider may reach $E_{pp}=100\,{\rm TeV}$. 
%We simply let $v$ be the number labeling energy from now on. 
Let $f_{q/p}(x_j,Q^2)$ be the PDF of quark $q_j$ inside a proton at the scale $Q$. So the BV cross section $\sigma_{\rm BV}(E_{pp})$ is given by, before the inclusion of the phase space factor,
%at total $pp$ energy $\sqrt{s}=2E$,
\begin{align} \label{eq:PDFBV}
\sigma_{\rm BV}(E_{pp}) & = \sum_{qq'}\int dx_1 f_{q/p}(x_1,s)\nonumber \\
& \qquad \times \int dx_2 f_{q'/p}(x_2,s){\hat\sigma}_{\rm BV}(v)\;,
\end{align}
where $E_{pp}\equiv\sqrt{s}$, and $v= \sqrt{x_1x_2} E_{pp}$.

\subsection{Crude estimate}
\label{sec:naive}

We can make a rough analytical estimate to get some idea, even though the resulting numerical values need improvement. This subsection estimates $\eta(E_{pp})$ using an unrealistic PDF and without consideration of band structure and phase space suppression. The analytical calculations below give us a general sense of what $\eta(E_{pp})$ looks like. 
As a start, we consider a simple (toy) PDF for valence quarks $q$ in a proton which is scale-independent, 
\begin{equation}
f _{q/p}(x)=A_q x^2 (1-x)^3\;,
\label{eq:pdf1}
\end{equation}
where $\int_0^1 dx f_{u/p}(x)=2$ and $\int_0^1 dx f_{d/p}(x)=1$.
This PDF (\ref{eq:pdf1}) allows an analytic discussion, but is only qualitatively valid. 

If we do not care about species, we shall choose $\int_0^1 dx f_{q/p}(x)=3$, so $A_q=180$.
The Bloch-wave picture indicates that the $\hat{\sigma}_{\rm BV}(v)$ is exponentially enhanced when $v \gtrsim E_{\rm sph}$ due to the overlap of high energy Bloch bands. Thus, for the purpose of estimating, we here simply take a cutoff model,
\begin{equation}
\hat{\sigma}_{\rm BV}(v) = 
\begin{cases}
\sigma_0\;, & x_1 x_2 > c \\
0 \;,& {\rm otherwise}
\end{cases}\;,
\label{eq:hat_sigma_simple}
\end{equation}
where $c=(E_{\rm sph}/E_{pp})^2$.
$\sigma_0$ is an overall normalization.
We may assume that, for $v \gg 10\,{\rm TeV}$, $\sigma_0\lesssim \sigma_{\rm total} (pp)$, where $\sigma_{\rm total} (pp)$ does not vary much.
Since we are comparing the BV event rate between different $E_{pp}$~\eqref{eq:eta},
%, we may normalize everything in the unit of $\sigma_{\rm BV}(13\,{\rm TeV})$. 
%Then, 
the value of $\sigma_0$ is not important here.
With this approximation, we could write
\begin{align}
\sigma_{\rm BV}(E_{pp}) & \approx  A_q^2 \sigma_0  \int_c^1 dx_1  (1-x_1)^3 \int_{c/x_1}^1 dx_2  (1-x_2)^3 \nonumber\\
 & = A_q^2 {\sigma}_0 \,G(c)\;,
\end{align}
where
\begin{align}
G(c) &=\frac{1}{3600} + \frac{10}{9} c^3 + \frac{27}{16} c^4 - \frac{54}{25} c^5 - \frac{23}{36} c^6 \nonumber\\
&\qquad +\left(\frac{1}{3} + \frac{9}{4}c + \frac{9}{5} c^2 + \frac{1}{6}c^3 \right) c^3 \ln c \nonumber\;.
\end{align}
As a check, we have $\sigma_{\rm BV}(c=1)=0$.

As a reasonable approximation, we take $E_{\rm sph} = 9 \,{\rm TeV}$ as a benchmark. For $E_{pp} = 13 \,{\rm TeV}$, $c=(9/13)^2=0.479$, while $c=0.413$ for $E_{pp}=14\,{\rm TeV}$, etc.

So we have 
    $\eta(13.6\,{\rm TeV}) = 1.80$ and $\eta(14\,{\rm TeV}) = 2.51$.
This indicates that only a factor of $2.5$ gains in going from $13 \,{\rm TeV}$ to $14 \,{\rm TeV}$. 
Compared to higher energies, 
we now have     $\eta(20\,{\rm TeV})=23.2$ and  $\eta(25\,{\rm TeV}) = 39.6$.
About a factor of $20$ gain from $13$ to $20\,{\rm TeV}$.
For even higher energies, $\eta(50\,{\rm TeV}) = 61.1$ and $\eta(100\,{\rm TeV}) = 62.8$.
One improves a little (1.03 gain) going from $50\,{\rm TeV}$ to $100\,{\rm TeV}$, which is much less efficient compared to the improvement from $13\,{\rm TeV}$ to $25\,{\rm TeV}$. This is due to the behavior at $x\to 0$ which comes from $x^2$ suppression. That is, the enhancement is saturated.

\subsection{Numerical estimate with $\theta$ phase space suppression}
\label{sec:theta}

Equation~\eqref{eq:hat_sigma_simple} is an oversimplification of the Bloch-wave solution. According to the Bloch-wave picture~\cite{Tye:2015tva,Tye:2017hfv}, we have $\hat{\sigma}_{\rm BV}(v) = \sigma_0$ if $v$ falls inside a Bloch wave band and $\hat{\sigma}_{\rm BV}(v) = 0$ otherwise. The center energies of Bloch bands and their widths are shown in Table~\ref{tab:bands}. ``Manton"~\cite{Manton:1983nd,Klinkhamer:1984di} and ``AKY"~\cite{Akiba:1988ay} refer to two different parametrizations of the Sphaleron potential. Here for those bands with energies above the first row in Table~\ref{tab:bands}, we consider them to be continuous due to the overlaps. So, for example, $\hat{\sigma}_{\rm BV}(v>9.113\,{\rm TeV})=\sigma_0$ for Manton potential. We neglect those bands with widths smaller than $10^{-9}\,{\rm TeV}$.

\begin{table}
    \centering
    \caption{Bloch wave bands from Ref.~\cite{Tye:2015tva}. Here $E_i$ is the band center energy and $\Delta_i$ is the band width. Those bands with a width smaller than $10^{-9}\,{\rm TeV}$ are neglected for they are essential zeros in our calculation precision, as explained in Sec.~\ref{sec:theta}.}
    \begin{ruledtabular}
    \begin{tabular}{cccc}
    \multicolumn{2}{c}{Manton} &    \multicolumn{2}{c}{AKY}  \\
    \cmidrule{1-2}
    \cmidrule{3-4}
    $E_i/{\rm TeV}$  & $\Delta_i/{\rm TeV}$  & $E_i/{\rm TeV}$  & $\Delta_i/{\rm TeV}$ \\
    $9.113$ & $0.01555$ & $9.110$ & $0.01134$ \\
    $9.081$ & $7.192\times10^{-3}$ & $9.084$ & $4.957\times10^{-3}$ \\
    $9.047$ & $2.621 \times 10^{-3}$ & $9.056$ & $1.718 \times 10^{-3}$ \\
    $9.010$ & $8.255 \times 10^{-4}$ & $9.026$ & $5.186 \times 10^{-4}$ \\
    $8.971$ & $2.382 \times 10^{-4}$ & $8.994$ & $1.438 \times 10^{-4}$ \\
    $8.931$ & $6.460 \times 10^{-5}$ & $8.961$ & $3.747 \times 10^{-5}$ \\
    $8.890$ & $1.666 \times 10^{-5}$ & $8.927$ & $9.279 \times 10^{-6}$ \\
    $8.847$ & $4.114 \times 10^{-6}$ & $8.892$ & $2.198 \times 10^{-6}$ \\
    $8.804$ & $9.779 \times 10^{-7}$ & $8.857$ & $5.008 \times 10^{-7}$ \\
    $8.759$ & $2.245 \times 10^{-7}$ & $8.802$ & $1.101 \times 10^{-7}$ \\
    $8.714$ & $4.993 \times 10^{-8}$ & $8.783$ & $2.341 \times 10^{-8}$ \\
    $8.668$ & $1.078 \times 10^{-8}$ & $8.745$ & $4.828 \times 10^{-9}$ \\
    $8.621$ & $2.262 \times 10^{-9}$ & & 
    \end{tabular}
    \end{ruledtabular}
    \label{tab:bands}
\end{table}

%CTEQ uses a more detailed form for valence quarks and gluons,
%\begin{equation}
%f _{q/p}(x)=Ax^{a_1-1}(1-x)^{a_2}e^{a_3x} [1+e^{a_{4}}x]^{a_5}
%\end{equation}

%To compare different $s$, we do not have to know the value of $A$. $f$ for quarks from the sea (i.e., anti-quarks) goes like $(1-x)^7$, so they may be ignored. ($f_g/p(x)$ for gluons go like $(1-x)^5$.)

\begin{figure}
\centering
\includegraphics[width=5.5cm]{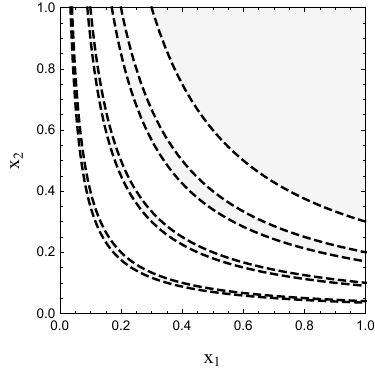}
\caption{A sketch of phase space for $x_1$--$x_2$. When considering the Bloch bands structure ~\cite{Tye:2015tva} in Eq.~\eqref{eq:PDFBV}, the integration over $x_1$--$x_2$ space should only include those satisfying Eq.~\eqref{eq:x1x2_band}, which has a pattern as the gray shaded region shown here.}
\label{fig:band_phase}
\end{figure}

%From the previous section, we have the rough picture that enhancement shall be significant going from $13\,{\rm TeV}$ to higher energy, and will be saturated when going beyond $50\,{\rm TeV}$. However, the PDF we used in the rough estimate Eq.~\eqref{eq:pdf1} is not precise and scale-independent. 
The PDF Eq.~\eqref{eq:pdf1} used in the last subsection is also too crude. Here we use realistic PDFs from the CTEQ program.
According to CT18~\cite{Hou:2019efy}, the PDFs at the initial scale $Q_0=1.3\,{\rm GeV}$ could be parametrized as
\[
    f_{q/p}(x,Q_0^2) = a_0 x^{a_1-1}(1-x)^{a_2} P_q(y_q;a_3,a_4,\cdots)\;,
\]
where $P_q(y_q;a_3,a_4,\cdots)$ and $y_q(x)$ are the polynomial functions that have different forms for each species. For PDFs at a higher energy scale, one could compute them by using renormalization equations. Details for those parameter values, polynomial forms, and higher scale evolution are included in Ref.~\cite{Hou:2019efy}.  Here we take the results from Ref.~\cite{Hou:2019efy} to estimate the $\eta(E_{pp})$. We extract PDFs from the {\tt CT18NNLO} dataset using {\tt LHAPDF}. To have a consistent precision, we take discrete values of PDFs with the step $\delta x=10^{-4}$. Thus, for all numerical results, we take only $4$ significant digits below.

%For illustration purposes, we show PDFs for quarks at two different scales in Fig.~\ref{fig:pdfs} to have a qualitative feeling about the enhancement. 
The PDFs morph for higher scale. $f(x \to 0,Q^2)$ will usually becomes larger for higher $Q$ for every species. Also, the contribution from sea quarks and valance quarks shall be comparable for small $x$. 
One should include more bands as collision energy goes higher,
%$c=(E_{\rm sph}/E_{pp})^2 = (9\,{\rm TeV})^2/s$ becomes smaller, 
and the integration region in $x_1$--$x_2$ phase space grows to include the smaller $x$ region 
%in Fig.~\ref{fig:simple_phase}. 
%This indicates that small $x$ shall contribute more to the BV process. 
This leads to the enhancement of the $\eta(E_{pp})$ for higher energies.

So far we have neglected the baryon-number conserving (BC) direction.  Recall that different $n$ states have different numbers of baryons and leptons and so their ground states have slightly different 
energies. The resulting effective sphaleron potential is a slightly tilted periodic potential. In quantum mechanics, this alone will suppress the BV process, i.e., $\Delta n=0$. It is the presence of the BC direction that allows finite $\Delta n$ BV process to happen~\cite{Qiu:2018wfb}. For our purpose here, we do not consider the tilted potential and take that including the BC direction in the phase space will further suppress the BV cross section. 

\begin{figure}
    \centering
    \includegraphics[width=6cm]{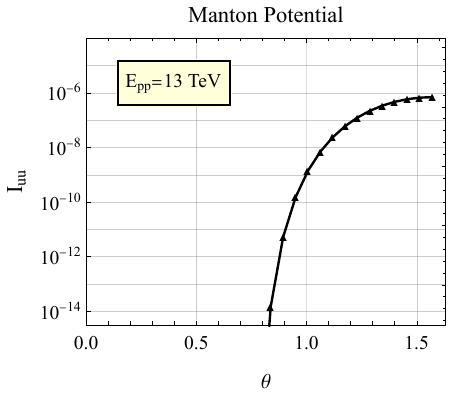}
    \caption{$I_{uu}$ as a function of $\theta$ for $E_{pp}=13\,{\rm TeV}$. Here Manton and AKY potentials lead to very similar results. Other $I_{qq'}$ also gives a similar suppression behavior.}
    \label{fig:Iuu}
\end{figure}

Here we consider a simple scenario, named $\theta$ phase space suppression (PSS). There are two orthogonal momentum directions in the phase space: the BC $\vec{p}_{\rm C}$ and BV $\vec{p}_{\rm V}$ directions. One can write down 
\begin{equation}
    \vec{p}_{qq}=\vec{p}_{\rm C} + \vec{p}_{\rm V}\;,\quad \vec{p}_{\rm C}\cdot \vec{p}_{\rm V}=0\;.
\end{equation}
In the relativistic limit, it could be converted to 
    $E_{qq}^2 \equiv v^2 = E_{\rm C}^2 + E_{\rm V}^2$,
where $E_{\rm C(V)}$ stands for the energy that goes into the baryon-number conserving (violating) direction. By introducing a parameter $\theta$, which is a random number that differs for every collision, one could conclude that only $E_{\rm V}=v\sin{\theta}$ shall participate in the BV process. Thus, the cross section is given by
\begin{align}
     \tilde{\sigma}_{\rm BV}(E_{pp},\theta) & = \sum_{q,q'} \int dx_1 f_{q/p}(x_1,s) \nonumber  \\
     &\qquad \times \int dx_2 f_{q'/p}(x_2,s) \hat{\sigma}_{\rm BV}(v\sin\theta) \nonumber \\
    & = \sigma_0 \sum_{q,q'} I_{qq'}(s,\theta)\;,
\end{align}
where
\begin{equation}
    I_{qq'}(s,\theta) = \int_{D(\theta)} dx_1 dx_2 f_{q/p}(x_1,s) f_{q'/p}(x_2,s)\;.
    \label{eq:Iqq'}
\end{equation}
Since we are considering Bloch bands here, such integration is performed over discontinuous bands as illustrated in Fig.~\ref{fig:band_phase}. 
Here $D(\theta)$ is the shaded region, 
\begin{equation}
    E_i-\frac{\Delta_i}{2}\leq\sqrt{x_1x_2}E_{pp}\sin{\theta} \leq E_i + \frac{\Delta_i}{2}
    \label{eq:x1x2_band}
\end{equation}
where 
$E_i$ is the center energy of $i$th Bloch band and $\Delta_i$ is its width. 
For fixed $E_{pp}$, one could see that smaller $\theta$ indicates that one has to integrate over lower bands region in the phase space, where the band gaps are relatively huge and widths are exponentially smaller.
Thus an extra suppression factor appears. 
Note that setting $\theta=\pi/2$ is equivalent to no suppression scenario.
As shown in Fig.~\ref{fig:Iuu}, smaller $\theta$ shall lead to huge suppression on the integration.
%As shown in Fig.~\ref{fig:Iuu}, the integration $I_{qq'}$ vanishes for $\theta< \arcsin{\sqrt{c}}$, which is due to the fact that $D(\theta)=0$ for such $\theta$.

The cross section $\tilde{\sigma}_{\rm BV}$ depends on $\theta$ for every event. We average out $\theta$ according to its probability density $P(\theta)$, to compare the efficiency for different $E_{pp}$ in observing BV events. Thus, we have
\begin{align}
    \sigma_{\rm BV}(E_{pp}) & = \int \tilde{\sigma}_{\rm BV}(E_{pp},\theta) P(\theta) d\theta \nonumber\\
    &= \frac{2}{\pi}\sigma_0 \sum_{q,q'}\int_0^{\pi/2}d \theta \,I_{qq'}(s,\theta) \;.
\end{align}
It is natural to assume that $\theta$ is sampled from a uniform distribution for every collision. Here we choose $P(\theta)=2/\pi$ for $\theta\in[0,\pi/2]$ in the second line above.

%We take realistic PDFs at the scale $Q=\sqrt{s}$ for a more realistic estimate. 
%Here
%Now the cross section is given by
%\begin{align}
%    \sigma_{\rm BV}(\sqrt{s}) & = \sigma_{\rm BV} \sum_{q,q'} \int_D dx_1 dx_2 \nonumber \\
%    & \qquad \times f_{q/p}(x_1,s) f_{q'/p}(x_2,s)\;,
%\end{align}
%where we take realistic PDFs at the scale $Q=\sqrt{s}$ for a more realistic estimate. Here
The summation runs over all quark species that participate in the BV process. For simplicity, we consider only the dominating contribution from
\begin{equation}
q,q'\in \{ u,d,s,\bar{u},\bar{d},\bar{s} \}\;.
\label{eq:qq'}
\end{equation}
The gluons do not participate in weak interactions and so contribute to the BV process only indirectly. So their contributions are not included here. 
%After numerical integration over the $x_1$--$x_2$ phase space, one obtains results as shown in Fig.~\ref{fig:sigma_s}. 

As a comparison between the simple cutoff [Eq.\eqref{eq:hat_sigma_simple}] and band structure model of $\hat{\sigma}_{\rm BV}(v)$, in Table~\ref{tab:sigma_comparision} we show the numerical result of $\sigma_{\rm BV}(E_{pp})/\sigma_0$ with various $E_{\rm sph}$ chosen for cutoff model together with the band structure.
Numbers in Table~\ref{tab:sigma_comparision} are all obtained under $\theta$ PSS for the sake of comparison.
%\st{Note that here they are both under the $\theta$ PSS.}
As one can see, the $\eta (E_{pp})$ result with the band structure is equivalent to a simple cutoff with an effective ${\hat E}_{\rm sph} \simeq 9.1\,{\rm TeV}$, slightly higher than the actual $E_{\rm sph}=9.0\,{\rm TeV}$. Also, one sees that the differences between Manton and AKY potentials are minor.

Figure~\ref{fig:sigma_s} shows the enhancement on $\sigma_{\rm BV}(E_{pp})$ in Manton potential. For comparison, in AKY potential, one finds $\eta (14 \,{\rm TeV}) \simeq 6.508 $, $\eta (20 \,{\rm TeV}) \simeq 1.842\times 10^3$ and $\eta (25 \,{\rm TeV}) \simeq 1.550\times 10^4$; that is a $4$ orders of magnitude enhancement going from $13\,{\rm TeV}$ to $25\,{\rm TeV}$. However, going from $50\,{\rm TeV}$ to $100\,{\rm TeV}$ will only give us roughly $1$ order-of-magnitude improvement in the event rate. Note that the size of phase space suppression from the random $\theta$ is about $1$ order of magnitude at the beginning, $E_{pp}\sim 13 \,{\rm TeV}$, and decreases to only roughly $0.5$ at $E_{pp} \sim 100\,{\rm TeV}$.

%A considerable enhancement (about 3 order-of-magnitude) shall be achieved when the energy is improved from $13\,{\rm TeV}$ to $20\,{\rm TeV}$. Meanwhile, going from $50\,{\rm TeV}$ to $100\,{\rm TeV}$ shall give us only about 1 order improvement, which indicates saturation. This is in agreement with our naive estimate in the last section.

\begin{table*}
    \centering
    \caption{$\sigma(E_{pp})$ with band structure and simple cutoff [Eq.~\eqref{eq:hat_sigma_simple})]. Here $\theta$ phase space suppression is applied. The first three columns are cutoff models and the last two are band models.}
    \begin{ruledtabular}   
    \begin{tabular}{l c c c c c  }
    \multirow{2}{*}{$E_{pp}/{\rm TeV}$}   & \multicolumn{5}{c}{$\sigma(E_{pp})/\sigma_0$}  \\
    \cmidrule{2-6}
     & $E_{\rm sph}=8.5\,{\rm TeV}$ & $E_{\rm sph}=9.0\,{\rm TeV}$ & $E_{\rm sph}=9.1\,{\rm TeV}$  & Manton & AKY \\ 
     \midrule
 $13$ & $8.106\times10^{-7}$ & $1.904\times10^{-7}$ & $1.398\times10^{-7}$ & $1.429\times10^{-7}$ & $1.414\times10^{-7}$ \\
 $13.6$ & $2.174\times10^{-6}$ & $6.013\times10^{-7}$ & $4.584\times10^{-7}$ & $4.670\times10^{-7}$ & $4.630\times10^{-7}$ \\
 $14$ & $3.881\times10^{-6}$ & $1.173\times10^{-6}$ & $9.119\times10^{-7}$ & $9.276\times10^{-7}$ & $9.203\times10^{-7}$ \\
 $20$ & $5.433\times10^{-4}$ & $2.940\times10^{-4}$ & $2.259\times10^{-4}$ & $2.615\times10^{-4}$ & $2.605\times10^{-4}$ \\
 $25$ & $0.003763$ & $0.002394$ & $0.002185$ & $0.002197$ & $0.002192$ \\
 $27$ & $0.006527$ & $0.004323$ & $0.003978$ & $0.003998$ & $0.003989$ \\
 $50$ & $0.1479$ & $0.1175$ & $0.1123$ & $0.1126$ & $0.1124$ \\
 $75$ & $0.5807$ & $0.4870$ & $0.4704$ & $0.4714$ & $0.4709$ \\
 $100$ & $1.264$ & $1.085$ & $1.053$ & $1.055$ & $1.054$ \\
    \end{tabular}
    \end{ruledtabular}
    \label{tab:sigma_comparision}
\end{table*}

\subsection{Numerical estimate with $K$ phase space suppression}

We consider another scenario, which simply introduces a suppression factor to the cross section integral, named $K$ phase space suppression. This is
\begin{align}
     \sigma_{\rm BV}(E_{pp}) &= \sigma_0 \sum_{q,q'}\int_{D} dx_1 dx_2 K(v)\nonumber\\
     & \qquad \qquad \times f_{q/p}(x_1,s) f_{q'/p}(x_2,s)\;,
\end{align}
where the integration $D=D(\pi/2)$ is the band structure consideration without $\theta$ suppression.

Naturally, the phase space suppression factor $K(v)$ shall interpolate from $0$ to $1$.
%satisfy $K(0)=0$ and $K(v\to\infty)=1$.
This is because when the energy is small, one shall expect little budget for BV. Meanwhile, when the energy is high enough, sphaleron potential could be neglected, and then the phase space suppression should vanish.
Also, $K(v)$ should be significantly enhanced when $v\sim E_{\rm sph}$ because the distinct scale in the BV process is $E_{\rm sph}$.
Thus, we assume that
\begin{equation}
    K(0)=0\;,\quad K(\infty)=1\;,\quad K(E_{\rm sph})\sim\mathcal{O}(0.1)\;.
    \label{eq:k_assumptions}
\end{equation}
Here we take a monotonically increasing function  
\begin{equation}
   K(v) =\left\{ \frac{2}{\pi} \arctan{\left[\left( \frac{v}{E_{\rm sph}}\right)^{\alpha}\right]}\right\}^{\beta}\;,
\end{equation}
which is parametrized by $\alpha>0$ and $\beta>0$. Note that $\beta=0$ corresponds to no suppression. 

\begin{figure}
    \centering
    \includegraphics[width=7.5cm]{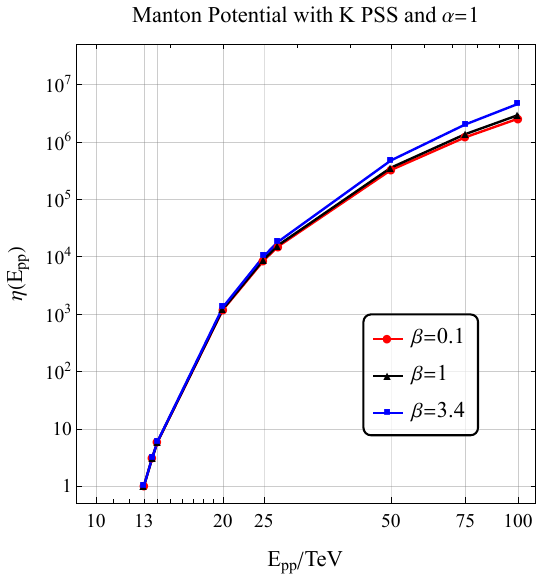}
    \caption{$\eta(E_{pp})$ with $K$ phase space suppression. Here we choose $E_{\rm sph}=9.0\,{\rm TeV}$ and $\alpha=1$.}
    \label{fig:eta_K_fixed_alpha}
\end{figure}

\begin{figure}
    \centering
    \includegraphics[width=7.5cm]{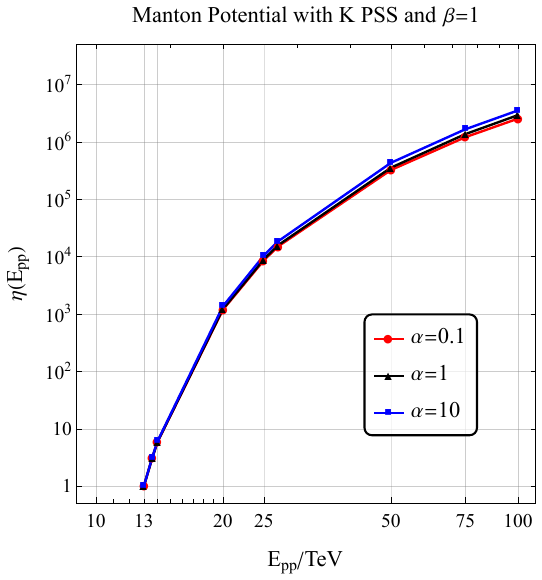}
    \caption{$\eta(E_{pp})$ with $K$ phase space suppression. Here we choose $E_{\rm sph}=9.0\,{\rm TeV}$ and $\beta=1$.}
    \label{fig:eta_K_fixed_beta}
\end{figure}

Adopting CT18 PDFs~\cite{Hou:2019efy} and considering quark content Eq.~\eqref{eq:qq'} in the Bloch band picture, we numerically calculate $\sigma_{\rm BV}(E_{pp})$ in unit of $\sigma_0$ with various choice of $\alpha$ and $\beta$ in Table~\ref{tab:sigma_K_manton} and~\ref{tab:sigma_K_aky}. Minor differences between Manton and AKY potentials are observed and order-of-magnitude behavior is the same. $K$ factor suppression is strong at low $E_{pp}$ and becomes weak when $E_{pp}$ go higher as anticipated.
Figure~\ref{fig:eta_K_fixed_alpha}
and Fig.~\ref{fig:eta_K_fixed_beta} show the enhancement factor $\eta(E_{pp})$ in the Manton potential, which is similar to the AKY potential. As shown in Fig.~\ref{fig:eta_K_fixed_beta}, varying $\alpha$ has little impact on $\eta(E_{pp})$. 
For larger $\beta>3.4$,
%\st{For $\beta=10$,}
one essentially changes the behavior of $K(E_{\rm sph})$, which shall lead to a significant change on $\eta(E_{pp})$ and against our assumption in Eq.~\eqref{eq:k_assumptions}. 
For reasonable choices of $\alpha$ and $\beta$, 
one shall have $\sim 4$ order enhancement on BV event rate going from $E_{pp}=13\,{\rm TeV}$ to $25\,{\rm TeV}$, and only about $1$ order gain from $E_{pp}=50\,{\rm TeV}$ to $100\,{\rm TeV}$.

\begin{table*}
    \centering
    \caption{$\sigma(E_{pp})$ with $K$ phase space suppression factor in band model of Manton potential. Here $E_{\rm sph}= 9\,{\rm TeV}$.}
    \begin{ruledtabular}   
    \begin{tabular}{l c c c c c c }
    \multirow{3}{*}{$E_{pp}/{\rm TeV}$}   & \multicolumn{6}{c}{$\sigma(E_{pp})/\sigma_0$}  \\
    \cmidrule{2-7}
     & $\beta=0$ & $\alpha=1$ & $\alpha=1$  & $\alpha=1$ & $\alpha=0.1$ & $\alpha=10$  \\ 
     & (No PSS) & $\beta=0.1$ & $\beta=1$  & $\beta=3.4$ & $\beta=1$ & $ \beta=1$  \\
     \midrule
      $13$ & $9.738\times10^{-7}$ & $9.111\times10^{-7}$ & $5.004\times10^{-7}$  & $1.014\times10^{-7}$ & $4.883\times10^{-7}$ &  $6.122\times10^{-7}$  \\
      $13.6$ &  $2.987\times10^{-6}$ & $2.796\times10^{-6}$ & $1.540\times10^{-6}$ & $3.142\times10^{-7}$ & $1.498\times10^{-6}$ &  $1.912\times10^{-6}$ \\
      $14$ & $5.716\times10^{-6}$ & $5.350\times10^{-6}$  & $2.951\times10^{-6}$ & $6.052\times10^{-7}$ & $2.867\times10^{-6}$ & $3.699\times10^{-6}$ \\
      $20$ & $0.001149$ & $0.001078$ & $ 6.081\times10^{-4}$ & $1.331\times 10^{-4}$ & $5.779\times 10^{-4}$ & $8.329\times 10^{-4}$ \\
      $25$ & $0.008322$ & $0.007821$ & $0.004481$ & $0.001028$ & $0.004193$ & $0.006363$ \\
      $27$ & $0.01448$ & $0.01362$ & $0.007846$ & $0.001831$ & $0.007302$ & $0.01125$ \\
      $50$ & $0.3106$ & $0.2936$ & $0.1773$ & $0.04803$ & $0.1576$ & $0.2637$ \\
      $75$ & $1.153$ & $1.093$ & $0.6807$ & $0.2034$ & $0.5874$ & $1.012$ \\
      $100$ & $2.424$ & $2.303$ & $1.461$ & $0.4647$ & $1.239$ & $2.159$ \\
    \end{tabular}
    \end{ruledtabular}
    \label{tab:sigma_K_manton}
\end{table*}

\begin{table*}
    \centering
    \caption{$\sigma(E_{pp})$ with $K$ phase space suppression factor in band model of AKY potential. Here $E_{\rm sph}= 9\,{\rm TeV}$.}
    \begin{ruledtabular}   
    \begin{tabular}{l c c c c c c }
    \multirow{3}{*}{$E_{pp}/{\rm TeV}$}   & \multicolumn{6}{c}{$\sigma(E_{pp})/\sigma_0$}  \\
    \cmidrule{2-7}
     & $\beta=0$ & $\alpha=1$ & $\alpha=1$  & $\alpha=1$ & $\alpha=0.1$ & $\alpha=10$  \\ 
     & (No PSS) & $\beta=0.1$ & $\beta=1$  & $\beta=3.4$ & $\beta=1$ & $ \beta=1$  \\
     \midrule
    $13$ & $9.648\times10^{-7}$ & $9.027\times10^{-7}$ & $4.959\times10^{-7}$ & $1.005\times10^{-7}$ & $4.838\times10^{-7}$ & $6.076\times10^{-7}$ \\
    $13.6$ & $2.963\times10^{-6}$ & $2.773\times10^{-6}$ & $1.527\times10^{-6}$ & $3.119\times10^{-7}$ & $1.486\times10^{-6}$ & $1.900\times10^{-6}$ \\
    $14$ & $5.673\times10^{-6}$ & $5.310\times10^{-6}$ & $2.930\times10^{-6}$ & $6.011\times10^{-7}$ & $2.846\times10^{-6}$ & $3.677\times10^{-6}$ \\
    $20$ & $0.001145$ & $0.001074$ & $6.061\times10^{-4}$ & $1.327\times10^{-4}$ & $5.759\times10^{-4}$ & $8.309\times10^{-4}$ \\
    $25$ & $0.008302$ & $0.007802$ & $0.004471$ & $0.001026$ & $0.004183$ & $0.006353$ \\
    $27$ & $0.01445$ & $0.01359$ & $0.007830$ & $0.001828$ & $0.007286$ & $0.01123$ \\
    $50$ & $0.3102$ & $0.2932$ & $0.1772$ & $0.04800$ & $0.1574$ & $0.2635$ \\
    $75$ & $1.152$ & $1.092$ & $0.6802$ & $0.2033$ & $0.5870$ & $1.011$ \\
    $100$ & $2.422$ & $2.301$ & $1.460$ & $0.4645$ & $1.238$ & $2.158$ 
    \end{tabular}
    \end{ruledtabular}
    \label{tab:sigma_K_aky}
\end{table*}

\section{Average Same Sign Charged Leptons per Event}

 Here $\eta$ (\ref{eq:eta}) only compares $\sigma_{\rm BV}$ at different energies. In reality, the initial $E_V$ is reduced as the CS number $Q$ (\ref{eq:QMeq}) moves $|\Delta n|$ steps, due to the production of $3|\Delta n|$ baryons and leptons. This lowering in energy will reduce the value of $|\Delta n|$ a BV scattering can reach. In the analysis of Ref.~\cite{Qiu:2018wfb}, we treat this effect as a tilt in the periodic sphaleron potential $V(Q)$ (\ref{eq:QMeq}). So we expect that $\sigma_{\rm BV}(25\, {\rm TeV})$ will involve events with larger $|\Delta n|$ than $\sigma_{\rm BV}(13\, {\rm TeV})$. In a single $\Delta n$ event, there are on average $3|\Delta n|/2$ same-sign charged leptons (and up to $3|\Delta n|$ same-sign charged leptons). A crude estimate suggests that the average $\langle \Delta n \rangle$ at 25 TeV is easily an order of magnitude bigger than the average $\langle \Delta n \rangle$ at 13 TeV. That is, the probability of BV detection can be $10^5$ higher at 25 TeV than at 13 TeV.
 
\section{Summary and Discussion}

In this short note, we demonstrate the enhancement of the baryon-number violating event rate when the COM energy for the $pp$ collider is increased. The estimate includes the Bloch band structure for unsuppressed BV scatterings and the phase space suppression from the baryon-number conserving direction. The Bloch band structure yields an effective cutoff of $E_{qq}\simeq 9.1$ TeV, a little above the simple cutoff of $E_{qq}\simeq 9.0$ TeV \footnote{Before turning on $U_{\rm Y}(1)$, $E_{\rm sph}=9.1$ TeV. Turning on $U_{\rm Y}(1)$ lowers it to $E_{\rm sph}=9.0$ TeV.}.  The phase space suppression factor is formulated in two ways, $\theta$ and $K$ phase space suppression. $\theta$ PSS scenario introduces a random parameter $\theta$ for every collision describing the energy budget of participating in the BV and BC process. We compare the event rate for different COM energy by integrating out $\theta$, which is sampled from a uniform distribution. $K$ PSS scenario introduces a monotonic function that describes the suppression from phase space. For reasonable choices of parameters in $K$, we have similar results as that in the $\theta$ PSS case. 
The precise values of $\eta(E_{pp})$ depend on the specific model (choice of the sphaleron potential and the phase space suppression factor). They are in general agreement with each other. Here, we treat these variations as uncertainties in $\eta(E_{pp})$. 

In summary, combining all scenarios considered above (except crude estimate in Sec.~\ref{sec:naive}), we now have ($\eta(13\,{\rm TeV})=1$ by definition), up to two significant digits,
\begin{align}
    \eta(13.6\,{\rm TeV})&\simeq 3.1\text{--}3.3 \;, \nonumber\\
    \eta(14\,{\rm TeV}) &\simeq 5.9\text{--}6.5 \;, \nonumber \\
    \eta(20\,{\rm TeV}) & \simeq 1.2\text{--}1.8\times 10^{3} \;, \nonumber\\
    \eta(25\,{\rm TeV}) & \simeq 0.86\text{--}1.6\times 10^{4} \;.
\end{align}
%About a factor of $3$ gain from $13$ to $20\,{\rm TeV}$.
For even higher energies, we have
\begin{align}
    \eta(50\,{\rm TeV}) &\simeq 3.2\text{--}7.9\times 10^{5}\;, \nonumber\\
    \eta(100\,{\rm TeV}) & \simeq 2.5\text{--}7.5\times 10^{6} \;.
\end{align}
The results indicate that increasing the COM $pp$ energy from $13\,{\rm TeV}$ to $25\,{\rm TeV}$ will yield a huge enhancement to the event rate. Together with the enhancement of $\langle \Delta n \rangle$ per event, the probability of BV detection can be $10^5$ higher at 25 TeV than at 13 TeV. Although the enhancement in $\sigma_{\rm BV}$ is more modest going from $50\,{\rm TeV}$ to $100\,{\rm TeV}$, the enhancement in $\langle \Delta n \rangle$ should be substantial.
 
\begin{acknowledgements}
We thank Sam Wong and Kirill Prokofiev for their useful discussions.
\end{acknowledgements} 

\bibliographystyle{utphys}
\bibliography{reference}

\providecommand{\href}[2]{#2}\begingroup\raggedright\begin{thebibliography}{10}

\bibitem{Belavin:1975fg}
A.~A. Belavin, A.~M. Polyakov, A.~S. Schwartz, and Y.~S. Tyupkin,
  ``{Pseudoparticle Solutions of the Yang-Mills Equations},''
  \href{http://dx.doi.org/10.1016/0370-2693(75)90163-X}{{\em Phys. Lett. B}
  {\bfseries 59} (1975) 85--87}.

\bibitem{tHooft:1976rip}
G.~'t~Hooft, ``{Symmetry Breaking Through Bell-Jackiw Anomalies},''
  \href{http://dx.doi.org/10.1103/PhysRevLett.37.8}{{\em Phys. Rev. Lett.}
  {\bfseries 37} (1976) 8--11}.

\bibitem{tHooft:1976snw}
G.~'t~Hooft, ``{Computation of the Quantum Effects Due to a Four-Dimensional
  Pseudoparticle},'' \href{http://dx.doi.org/10.1103/PhysRevD.14.3432}{{\em
  Phys. Rev. D} {\bfseries 14} (1976) 3432--3450}. [Erratum: Phys.Rev.D 18,
  2199 (1978)].

\bibitem{Tye:2015tva}
S.~H.~H. Tye and S.~S.~C. Wong, ``{Bloch Wave Function for the Periodic
  Sphaleron Potential and Unsuppressed Baryon and Lepton Number Violating
  Processes},'' \href{http://dx.doi.org/10.1103/PhysRevD.92.045005}{{\em Phys.
  Rev. D} {\bfseries 92} (2015) 045005},
  \href{http://arxiv.org/abs/1505.03690}{{\ttfamily arXiv:1505.03690
  [hep-th]}}.

\bibitem{Manton:1983nd}
N.~S. Manton, ``{Topology in the Weinberg-Salam Theory},''
  \href{http://dx.doi.org/10.1103/PhysRevD.28.2019}{{\em Phys. Rev. D}
  {\bfseries 28} (1983) 2019}.

\bibitem{Klinkhamer:1984di}
F.~R. Klinkhamer and N.~S. Manton, ``{A Saddle Point Solution in the
  Weinberg-Salam Theory},''
  \href{http://dx.doi.org/10.1103/PhysRevD.30.2212}{{\em Phys. Rev. D}
  {\bfseries 30} (1984) 2212}.

\bibitem{Akiba:1988ay}
T.~Akiba, H.~Kikuchi, and T.~Yanagida, ``{Static Minimum Energy Path From a
  Vacuum to a Sphaleron in the {Weinberg-Salam} Model},''
  \href{http://dx.doi.org/10.1103/PhysRevD.38.1937}{{\em Phys. Rev. D}
  {\bfseries 38} (1988) 1937--1941}.

\bibitem{Bachas:2016ffl}
C.~Bachas and T.~Tomaras, ``{Band Structure in Yang-Mills Theories},''
  \href{http://dx.doi.org/10.1007/JHEP05(2016)143}{{\em JHEP} {\bfseries 05}
  (2016) 143}, \href{http://arxiv.org/abs/1603.08749}{{\ttfamily
  arXiv:1603.08749 [hep-th]}}.

\bibitem{Ellis:2016ast}
J.~Ellis and K.~Sakurai, ``{Search for Sphalerons in Proton-Proton
  Collisions},'' \href{http://dx.doi.org/10.1007/JHEP04(2016)086}{{\em JHEP}
  {\bfseries 04} (2016) 086}, \href{http://arxiv.org/abs/1601.03654}{{\ttfamily
  arXiv:1601.03654 [hep-ph]}}.

\bibitem{Hou:2019efy}
T.-J. Hou {\em et~al.}, ``{New CTEQ global analysis of quantum chromodynamics
  with high-precision data from the LHC},''
  \href{http://dx.doi.org/10.1103/PhysRevD.103.014013}{{\em Phys. Rev. D}
  {\bfseries 103} (2021) 014013},
  \href{http://arxiv.org/abs/1912.10053}{{\ttfamily arXiv:1912.10053
  [hep-ph]}}.

\bibitem{Qiu:2018wfb}
Y.-C. Qiu and S.~H.~H. Tye, ``{Role of Bloch Waves in baryon-number violating
  processes},'' \href{http://dx.doi.org/10.1103/PhysRevD.100.033006}{{\em Phys.
  Rev. D} {\bfseries 100} (2019) 033006},
  \href{http://arxiv.org/abs/1812.07181}{{\ttfamily arXiv:1812.07181
  [hep-ph]}}.

\bibitem{Tye:2017hfv}
S.~H.~H. Tye and S.~S.~C. Wong, ``{Baryon Number Violating Scatterings in
  Laboratories},'' \href{http://dx.doi.org/10.1103/PhysRevD.96.093004}{{\em
  Phys. Rev. D} {\bfseries 96} (2017) 093004},
  \href{http://arxiv.org/abs/1710.07223}{{\ttfamily arXiv:1710.07223
  [hep-ph]}}.

\end{thebibliography}\endgroup

\end{document}